\documentclass[]{article}

\usepackage[]{geometry}

\usepackage{amsmath,amsthm} 
\usepackage{amssymb}  
\usepackage{amsfonts}
\usepackage{graphicx}
\usepackage{epstopdf}
\usepackage{mathtools}
\DeclarePairedDelimiter\ceil{\lceil}{\rceil}

\usepackage{color,hyperref}
\usepackage{cite}
\makeatletter

\newtheorem{theorem}{Theorem}[section]
\newtheorem{corollary}[theorem]{Corollary}
\newtheorem{lemma}[theorem]{Lemma}
\newtheorem{remark}[theorem]{Remark}

\newtheorem{definition}[theorem]{Definition} 
\newtheorem{conjecture}[theorem]{Conjecture} 

\newcommand{\TheTitle}{On Asymptotic Log-Optimal Buy-and-Hold Strategy}

\title{{\TheTitle}
}

\author{
	Chung-Han Hsieh\thanks{
		  Department of Quantitative Finance, National Tsing Hua University, Hsinchu 300044, Taiwan R.O.C.,
		E--mail: \mbox{\href{mailto:ch.hsieh@mx.nthu.edu.tw}{ch.hsieh@mx.nthu.edu.tw}}.}
}
\date{}
\begin{document}

\maketitle

\begin{abstract}
In this paper, we consider a frequency-based portfolio optimization problem with $m \geq 2$ assets when the expected logarithmic growth (ELG) rate of wealth is used as the performance metric. 
With the aid of the notion called \textit{dominant asset}, it is known that the optimal ELG level is achieved by investing all available funds on that asset.
However, such an ``all-in" strategy is arguably too risky to implement in practice. 
Motivated by this issue, we study the case where the portfolio weights are chosen in a rather \textit{ad-hoc} manner and a buy-and-hold strategy is subsequently used.
Then we show that, if the underlying portfolio contains a dominant asset, buy and hold on that specific asset is asymptotically log-optimal with a sublinear rate of convergence. 
This result also extends to the scenario where a trader either does not have a probabilistic model for the returns or does not trust a model obtained from historical data. 
To be more specific, we show that if a market contains a dominant asset, buy and hold a \textit{market} portfolio involving nonzero weights for each asset is asymptotically log-optimal. 
Additionally, this paper also includes a conjecture regarding the property called \textit{high-frequency maximality}. That is, in the absence of transaction costs,   high-frequency rebalancing is \textit{unbeatable} in the ELG sense. Support for the conjecture, involving a lemma for a weak version of the conjecture,  is provided. 
This conjecture, if true, enables us to  improve the log-optimality result obtained previously. Finally, a result that indicates a way regarding an issue about \textit{when} should one to rebalance their portfolio, if needed, is also provided.
Examples, some involving simulations with historical data, are also provided along the way to illustrate the~theory.
\end{abstract}



\section{Introduction}\label{SECTION: Introduction}
For a large class of multiperiod memoryless betting games, \cite{Kelly_1956} in his seminal paper proposed an approach which calls for maximizing the Expected Logarithmic Growth~(ELG) of a bettor's wealth. 
This ELG maximization approach, the so-called Kelly Criterion, has indeed resulted in a voluminous body of literature with
various theoretical generalizations and ramifications; e.g., see \cite{ algoet1988asymptotic} for a study on asymptotic properties of ELG optimal investment, \cite{bell1988game} for study an ELG approach in game-theoretic setting, \cite{hakansson1971optimal,thorp1975portfolio, cover1991universal, rotando1992kelly, maslov1998optimal} for asset management and stock trading, \cite{thorp2006kelly} for applications in Blackjack and sports betting, 
\cite{rujeerapaiboon2016robust} for a robust ELG formulation,  and~\cite{cover2006elements,Luenberger_2011} for a good summary of ELG approach. See also~\cite{maclean2011kelly} for an extensive survey covering many of the most important papers. 
A sampling of more recent papers on the ELG topics
includes~\cite{hsieh2016kelly,lo2018growth,hsieh2018frequency,hsieh2018rebalancing,wu2020analysis, hsieh2021necessary,obrien2021generalization}.

\smallskip
Compared with many of the recent papers that contributed to the ELG maximization problem and its application to stock trading, the effects of \textit{rebalancing frequency} are arguably \textit{not} heavily considered into the existing~literature. 
Some initial results along this line regarding rebalancing frequency effects can be found in~\cite{kuhn2010analysis, das2014computing, das2015computing, hsieh2018frequency,hsieh2018rebalancing,hsieh2021necessary}. 
Specifically, in \cite{kuhn2010analysis}, a portfolio optimization with returns following a continuous geometric Brownian motion was considered. 
It then indicated that continuous rebalancing slightly outperforms discrete rebalancing if there are no transaction costs and if the rebalancing intervals are short enough. 
On the other hand, in~\cite{das2014computing, das2015computing},  a portfolio optimization was considered with the constant weight~$K$ selected \textit{without} regard for the frequency with which the portfolio rebalancing is done. 
Subsequently, when this same weight~$K$ is used to find an optimal rebalancing period, the resulting levels of ELG are arguably suboptimal. 

\smallskip
In contrast to~\cite{kuhn2010analysis} and \cite{das2014computing,das2015computing}, we consider a general discrete-time portfolio optimization formulation using ELG as performance metric and seek an ``log-optimal" portfolio which depends on the rebalancing frequency.  
Our formulation to follow considers the entire range of rebalancing frequencies and both the probability distribution of the returns and the time interval between rebalances are arbitrary.

\medskip
\subsection{Plan for the Remainder of the Paper}
In Section~\ref{SECTION: Preliminaries}, we first provide some preliminaries  that are needed for the rest of this paper. Specifically, we formulate a frequency-based portfolio optimization problem using expected logarithmic growth as the performance metric and two technical notions called \textit{relative attractiveness} and \textit{dominant asset} are also reviewed. 
Then, in Section~\ref{SECTION: Main Results}, with the aid of dominant asset, we show that the ELG level obtained by a buy-and-hold strategy can asymptotically tend to the optimal~ELG level with a sublinear rate of convergence. 
Subsequently, in Section~\ref{Improvement of the Asymptotic Log-Optimality Theorem}, we provide a conjecture regarding the high-frequency maximality, which says that high-frequency rebalancing is \textit{unbeatable} in the ELG sense. 
Support of the conjecture, involving a lemma for a special case of the conjecture, is provided.  
Then, we indicate that the conjecture, if true, can be used to improve the prior asymptotic log-optimality result.
Examples, some involving simulations with historical data, are also provided along the way to illustrate the theory.
In Section~\ref{SECTION: Conclusion}, some concluding remarks and possible future research directions are~provided. Finally, in Appendix, a technical proof for the lemma regarding high-frequency maximality is provided.

\bigskip
\section{Preliminaries}
\label{SECTION: Preliminaries}
To study the effects of rebalancing frequency, we let $\Delta t > 0$ be the minimum time interval that is available for rebalancing trades. For example, in a typical high-frequency trading scenario, $\Delta t$ can be as small as a fraction of a second; see also an example in Section~\ref{Subsection:Illustrative Example Using Historical Intraday Tick Data}. Now with $n\geq 1$ being an integer, the \textit{rebalancing frequency} is defined as~$f:=1/(n\Delta t)$; i.e., one rebalances its portfolio per $n\Delta t$ unit of time. In the sequel, it is convenient to simply call that the integer $n$ to be the \textit{rebalancing period}. In this setting, the high-frequency rebalancing corresponds to the case when $n := 1$; i.e., one rebalances per  $\Delta t$ unit of time. Similarly, relatively low-frequency rebalancing corresponds to the case where $n > 1$; i.e., one waits for extra $n$ steps before rebalances.

%

\smallskip 
Now for~$k=0,1,\dots,$ we consider a trader who is forming a portfolio consisting of~$m \geq 2$ assets and assume that at least one of them is riskless with nonnegative rate of return $r \geq 0$, which is deterministic and is treated as a degenerate random variable with value $r$ for all $k$ with probability one. 
Alternatively, if Asset~$i$ is a stock whose price at stage~$k$ is~$S_i(k)>0$, then its \textit{return} is given~by
$$
X_i(k) = \frac{S_i(k+1) - S_i(k)}{S_i(k)}.
$$
In the sequel, for stocks, we assume that the return vectors
$
X(k):=\left[X_1(k) \, X_2(k)\,  \cdots \,X_m(k)\right]^T
$
have a \textit{known} distribution and have components $X_i(\cdot)$ which can be arbitrarily correlated.\footnote{Again, if the $i$th asset is riskless, then we put $X_i(k) = r \geq 0$ with probability one. If a trader maintains \textit{cash} in its portfolio, then this corresponds to the case~\mbox{$r=0.$}}
We also assume that these vectors $\{X(k): k\geq 0\}$ are independent and identically distributed (i.i.d.) with components satisfying
$
X_{\min,i}  \leq X_i(k) \leq X_{\max,i}
$
with known bounds above and with $X_{\max,i}$ being finite and~\mbox{$X_{\min,i} > -1$} for $i=1,2,\dots,m$. 
The latter constraint on $X_{\min,i}$ means that the loss per time step is limited to less than~$100\%$ and the price of a stock cannot drop to~zero.

\medskip
\subsection{Idealized Market}
We further assume that stock trading occurs within an ``idealized market." That is, we assume zero interest rates, zero transaction costs, and perfect liquidity conditions. There is no bid-ask spread, and the trader can buy
or sell any number of shares including fractions at the traded price $S_i(k)$. These assumptions arise in the finance literature in the context of ``frictionless" market; see~\cite{merton1992continuous}.

\medskip
\subsection{Trading Rules and Account Dynamics} 
Given initial account value $V(0)>0$, the trader fixes a rebalancing period~$n\geq 1$  and places an initial investment for each $i$th asset described by~$u_i(0):= K_iV(0)$ with $K_i \geq 0$ for $i=1,2,\dots, m$. Equivalently, this investment can be represented in terms of 
$
N_i(0)  := {u_i(0)}/{S_i(0)} 
$
shares.\footnote{The weight $K_i\geq 0$ disallows \textit{short selling} and the associated trades are going \textit{long}; i.e.,  the trader purchases shares from the broker in the hope of making a profit from a subsequent rise in the price of the underlying stock.}
Then, the corresponding account value at stage~$n$ is described by
\begin{align*}
	V(n) &= V(0) +  \sum_{i=1}^m N_i(0)(S_i(n) - S_i(0))\\[1ex]
	&= V(0) + \sum_{i=1}^m K_iV(0) \bigg( \frac{S_i(n)}{S_i(0)}- 1\bigg)\\[1ex]
	&= V(0) + V(0) \sum_{i=1}^m K_i  \bigg[ \prod_{k = 0}^{n-1}(1+X_i(k))- 1\bigg]\\[1ex]
	&= (1+ K^T \mathcal{X}_n) V(0)
\end{align*}
where $K:=[K_1\,\, K_2\,\, \cdots\,\, K_m]^T$ is the \textit{portfolio weight} vector satisfying some certain trading constraints which will be described in the next Subsection~\ref{subsection: Admissible Set of Portfolio Weights} and $\mathcal{X}_n := \left[\mathcal{X}_1(k) \,\, \mathcal{X}_2(k)\,  \cdots \,\mathcal{X}_m(k)\right]^T$~with
\[
\mathcal{X}_{n,i} := \prod_{k = 0}^{n-1}(1+X_i(k))- 1
\] being
the \textit{compound return} with known~bounds $\mathcal{X}_{n,i}>-1$ for all $n\geq 1$ and $i=1,2,\dots,m$. 
In the sequel, we may sometimes write $V(n,K)$ instead of $V(n)$ to emphasize the dependence on the portfolio weight~$K$.

\medskip
\subsection{Admissible Set of Portfolio Weights}\label{subsection: Admissible Set of Portfolio Weights}
Since~$K_i \geq 0$ for all $i=1,2,\dots,m$, the trade is going long. Henceforth, we consider the unit simplex constraint for the portfolio weights; i.e.,
$$
K \in {\mathcal K} := \left\{K \in \mathbb{R}^{m}: K_i \geq 0 \text{ for all $i$}, \; \sum_{i=1}^m K_i = 1 \right\}
$$ 
which is classical in finance; e.g., see \cite{cover1991universal,cover2006elements,Luenberger_2011,  hsieh2018rebalancing}.
That is, with~\mbox{$K \in \mathcal K$}, we have a guarantee that~100\% of the account is invested. Moreover, we note that the constraint set $\mathcal{K}$ assures trader's survivability; i.e., no bankruptcy is guaranteed; see also~\cite{hsieh2021necessary} for a  discussion of this important property.

\medskip
\subsection[Frequency-Dependent Optimization Problem]{Frequency-Dependent Optimization Problem} 
Consistent with the existing work in \cite{hsieh2018frequency,hsieh2018rebalancing}, for any rebalancing period~$n \geq 1$, we 
study the problem of maximizing the expected logarithmic~growth 
\begin{align*}
	{g_n}(K) 
	&:=  \frac{1}{n}\mathbb{E}\left[ \log \frac{V(n,K)}{V(0)} \right] \\[1ex]
	&= \frac{1}{n}\mathbb{E}\left[ {\log (1 + {K^T}{\mathcal{X}_n})} \right]
\end{align*}
and we use $g_n^*$ to denote the associated optimal expected logarithmic~growth.\footnote{It is readily verified that $g_n(K)$ is concave in~$K$ and the constraint set $\mathcal{K}$ is convex. Hence, while there is often no closed-form solution to the ELG maximization problem, it can be solved by a convex programming technique in a rather efficient way; e.g., see~\cite{boyd2004convex} for a detailed discussion on this topic.}
Furthermore, any vector~{$K_n^* \in \mathcal{K}$} satisfying~$g_n(K^*) = g_n^*$ is called a \textit{log-optimal weight with rebalancing period $n$}. When it is clear from the context, in the sequel, we may drop the subscript $n$ and simply write $K^*$ instead of $K_n^*.$
The portfolio which uses the log-optimal weight is called \textit{log-optimal portfolio}. Henceforth, whenever convenient, we may refer that the trader who adopts the log-optimal portfolio as a \textit{log-optimal trader}.  

%
%
%
%
%

\medskip
\subsection[Notion of Relative Attractiveness and Dominant Asset]{Notion of Relative Attractiveness and Dominant Asset} 
In this subsection, we review two technical notions called \textit{relative attractiveness} and \textit{dominant~asset} that are useful for the later development. Indeed, according to \cite{hsieh2018rebalancing,hsieh2021necessary}, these two notions enable us to state a result called the Dominate Asset Theorem, which tells us that if a portfolio contains a dominant asset, then a log-optimal trader should invest all available funds on that asset.

\medskip
\begin{definition}[Relative Attractiveness and Dominant Asset]
	{\rm Given a collection of $m\geq 2$ assets, we say that Asset $j$ is \textit{relatively more attractive than} Asset $i$ if 
		\begin{align}\label{def: relative attractiveness}
			\mathbb{E}\left[\frac{1+X_i(0)}{1+X_j(0)}\right]\leq 1.
		\end{align} 
		Asset $j$ is said to be  \textit{dominant}  if it is relatively more attractive than every other Asset $i \neq j$.}
\end{definition}

\medskip
\begin{remark} \rm The notion of relative attractiveness can be viewed as a measure of ``favorableness" to a long-only trader. For instance, a riskless Asset $j$ with nonnegative rate of returns $r\geq 0$ is relatively more attractive than Asset $i$ if and only~if
	\[
	\mathbb{E}\left[ \frac{1+X_i(0)}{1+r}\right] \leq 1.
	\]
	Equivalently, this implies that
	$
	r \geq \mathbb{E}\left[ {X_i(0)}\right] 
	$, which is consistent with the intuition that the Asset~$j$ may be more favorable to a long trader.
	We are now ready to state an important result derived from the notion of relative attractiveness and dominant asset.
\end{remark}

\medskip
\begin{theorem}[Dominant Asset]\label{thm: Dominant Asset} For the frequency dependent optimization problem defined above, 
	the log-optimal portfolio weight $K^*=e_j$ if and only if the Asset~$j$ is a dominant asset; i.e.,
	\[
	\mathbb{E}\left[ \frac{1+X_i(0)}{1+X_j(0)}\right] \leq 1.
	\]
	Furthermore, the resulting optimal expected logarithmic
	growth rate is given by
	$$
	g_n^* = g_1^* = \mathbb{E}[\log(1+X_j(0)].
	$$
\end{theorem}

\medskip
\proof{Proof.}
The detailed proof can be found in~\cite{hsieh2018rebalancing} and \cite{hsieh2021necessary}; hence is~omitted. \qedhere
\endproof

\medskip
\begin{remark} \rm In the theorem above, we see that the optimal ELG level is identical to $g_1^*$.  In the sequel, we shall use it as our performance benchmark.
\end{remark}

\bigskip
\section{Asymptotic Log-Optimality Theorem}
\label{SECTION: Main Results}
It is arguable that investing all available funds on a \textit{dominant} asset, while is log-optimal, is rather risky in practice. This issue motivates us to study the ELG performance using only a buy-and-hold strategy with ad-hoc weights. 
In this section,
With the aid of a dominant asset, we show that such a buy-and-hold strategy can be  asymptotic log-optimal. 
This is summarized in the next~theorem.

\medskip
\begin{theorem}[Asymptotic Log-Optimality of Buy and Hold] \label{thm: Second Asymptotic Log-Optimality of Buy and Hold}
	Let Asset~$j$ be a dominant asset for some $j\in \{1,2,\dots,m\}$. Given any $K \in \mathcal{K}$ with $K_j >0$, then 
	$
	\lim_{n \to \infty}g_n(K)  = g_1^*
	$
	with rate of convergence above governed~by  
	\[
	\frac{1}{n} \left( \log \dfrac {1}{K_j} +1 -\frac{1}{K_j} \right) \leq g_1^*-{g_n}\left( K \right) \leq \frac{1}{n}\left( \log \frac{1}{K_j} \right).
	\]
\end{theorem}

\medskip
\begin{proof}{Proof.} We begin by noting that if $K_j=1$, then we are back into the world of Dominant Asset Theorem and $g_1^*-g_n(K) =0$. Hence, we assume $K_j \in (0,1)$.
	Now to prove the upper bound for the performance difference $g_1^*-g_n(K)$. For notational convenience, we work with the total return vector 
	$\mathcal{R}_n =\mathcal{X}_n + \textbf{1}
	$ with $i$th component $\mathcal{R}_{n,i} = \mathcal{X}_{n,i}+1$ and $\textbf{1} =[1 \;\; 1 \cdots \; 1]^T \in \mathbb{R}^m$. 
	Note that $\mathcal{R}_{n,i}>0$ for all $i=1,\dots,m$ and all $n\geq 1$. Then, for $K\in \mathcal{K}$ with $K_j \in (0,1)$, we observe~that
	\begin{align}\label{eq: g_n}
		{g_n}\left( K \right) 
		&= \frac{1}{n}\mathbb{E}\left[ {\log  {K^T}{\mathcal{R}_n} } \right] \nonumber \\[1ex]
		&= \frac{1}{n}\mathbb{E}\left[ \log  \left( \sum_{i=1}^m K_i \mathcal{R}_{n,i} \right) \right] \nonumber \\[1ex]
		&= \frac{1}{n}\mathbb{E}\left[ {\log \left( {{K_j}{\mathcal{R}_{n,j}}\left( {1 + \sum\limits_{i \ne j} {\frac{{{K_i}{\mathcal{R}_{n,i}}}}{{{K_j}{\mathcal{R}_{n,j}}}}} } \right)} \right)} \right] \nonumber \\[1ex]
		&= \frac{1}{n}\log {K_j} + \frac{1}{n}\mathbb{E}\left[ \log  \mathcal{R}_{n,j}  \right] 
		+ \frac{1}{n}\mathbb{E}\left[ {\log \left( {1 + \sum\limits_{i \ne j} {\frac{{{K_i}{\mathcal{R}_{n,i}}}}{{{K_j}{\mathcal{R}_{n,j}}}}} } \right)} \right].
	\end{align} 
	Using the definition of $\mathcal{R}_{n,j}$, we note that the middle term on the equation~(\ref{eq: g_n}) is
	\begin{align*}
		\frac{1}{n} \mathbb{E}[\log \mathcal{R}_{n,j} ]&=\frac{1}{n} \mathbb{E}[\log (1+\mathcal{X}_{n,j}) ]\\[1ex]
		& =\frac{1}{n} \mathbb{E}\left[\log\prod_{k=0}^{n-1}(1+X_j(k)) \right]\\[1ex]
		&= \mathbb{E}\left[  \log(1+X_j(0)\right]\\[1ex]
		&= g_1^*
	\end{align*}
	where the second last equality holds because $X_i(k)$ are i.i.d. in $k$ and the last equality holds because Asset~$j$ is dominate and hence $g_1^*$ is achieved by Dominant Asset Theorem~\ref{thm: Dominant Asset}.
	Therefore, equation~(\ref{eq: g_n}) can be~rewritten as
	\begin{align}
		g_1^* -  {g_n}\left( K \right)  &= \frac{1}{n}\log {\frac{1}{K_j}} 
		- \frac{1}{n}\mathbb{E}\left[ {\log \left( {1 + \sum_{i \ne j} {\frac{{{K_i}{\mathcal{R}_{n,i}}}}{{{K_j}{\mathcal{R}_{n,j}}}}} } \right)} \right] \label{eq: g_1 minus gn}\\
		&\leq \frac{1}{n}\log \frac{1}{K_j} \nonumber
	\end{align} 
	where the last inequality holds since
	$
	\mathbb{E}\left[ {\log \left( {1 + \sum_{i \ne j} {\frac{{{K_i}{\mathcal{R}_{n,i}}}}{{{K_j}{\mathcal{R}_{n,j}}}}} } \right)} \right]\geq 0.
	$
	To prove the lower bound, we use the fact that $\log(1+z) \leq z$ for all $z>-1$. Then, it follows that
	\begin{align*}
		g_1^* -	{g_n}\left( K \right)
		&\geq \frac{1}{n}\log \dfrac 1 K_j - \frac{1}{n}\mathbb{E}\left[ \sum\limits_{i \ne j} {\frac{{{K_i}{\mathcal{R}_{n,i}}}}{{{K_j}{\mathcal{R}_{n,j}}}}}  \right]\\[1ex]
		&\geq \frac{1}{n}\log \dfrac 1 K_j - \frac{1}{n} \sum\limits_{i \ne j}\frac{K_i}{K_j}\mathbb{E}\left[ {\frac{\mathcal{R}_{n,i}}{\mathcal{R}_{n,j}}}  \right]\\[1ex]
		&\geq \frac{1}{n}\log \dfrac 1 K_j - \frac{1}{n} \sum\limits_{i \ne j}\frac{K_i}{K_j}\mathbb{E}\left[\prod_{k=0}^{n-1} {\frac{1+X_i(k)}{1+X_j(k)}}  \right].
	\end{align*}
	With the aid that the return sequence $\{X_i(k): k\geq 0\}$ are i.i.d., we have
	\begin{align*}
		g_1^* -	{g_n}\left( K \right) \geq \frac{1}{n}\log \dfrac 1 K_j - \frac{1}{n} \sum\limits_{i \ne j}\frac{K_i}{K_j} \left( \mathbb{E}\left[{\frac{1+X_i(0)}{1+X_j(0)}} \right] \right)^n. 
	\end{align*}
	Since Asset~$j$ is dominant, by Definition~\ref{def: relative attractiveness}, we have $\mathbb{E}\left[{\frac{1+X_i(0)}{1+X_j(0)}} \right] \leq 1$. Hence, a straightforward calculation leads to
	\begin{align*}
		g_1^* -	{g_n}\left( K \right) &\geq \frac{1}{n}\log \dfrac 1 K_j - \frac{1}{n} \sum\limits_{i \ne j}\frac{K_i}{K_j} \\
		&=\frac{1}{n} \left( \log \dfrac 1 K_j +1 -\frac{1}{K_j} \right).
	\end{align*}
	Lastly, it remains to verify that $\frac{1}{n} \left( \log \frac{1}{K_j} +1 -\frac{1}{K_j} \right) \leq \frac{1}{n}\left( \log \frac{1}{K_j} \right)$, which is equivalent to check if $  1 -\frac{1}{K_j}  \leq  0$. To see this, using the fact that $K_j \in (0,1]$, the desired inequality follows immediately. 
	To complete the proof, we note that both of the lower and upper bounds converge to zero as~$n\to\infty$. By Squeeze Theorem, it implies that $g_n(K)\to g_1^*$ as $n\to \infty$, which is desired. \qedhere
\end{proof}

\medskip
\begin{remark}\rm
 Theorem~\ref{thm: Second Asymptotic Log-Optimality of Buy and Hold} is practically useful in the following sense: If a dominant asset exists but is unknown to the trader, then investing all available fund in the dominant asset, while is log-optimal,  is \textit{not} needed in the long term. One can simply specify  nonzero weights on each asset and the desired ``asymptotic" log-optimality follows. 
	An interesting ramification is that if a trader buy and hold on a well-diversified portfolio involving one asset which is dominant, then asymptotic log-optimality follows \textit{automatically}. This is summarized in the following~corollary.
\end{remark}

\medskip
\begin{corollary}
	If a stock market contains a dominant asset. Then buy and hold a corresponding market portfolio with nonzero weights for each asset is asymptotically~log-optimal. 
\end{corollary}

\medskip
\begin{proof}{Proof.} The proof is a  trivial observation.
	Suppose the market has $m\geq 2$ assets and let Asset~$j$ be a dominant asset for some $j \in \{1,\dots,m\}$. 
	Now, we construct a market portfolio with the  weights satisfying $K_i>0$ for all $i=1,\dots,m$ and $\sum_i K_i = 1$. Such a portfolio is easy to construct, for example, one can simply take $K_i = 1/m$ for all $i$.
	Then, $K \in \mathcal{K}$. More importantly, the nonzero weights assumption assures that the weight for Asset~$j$ is nonzero; i.e., $K_j >0.$
	Now, applying Theorem~\ref{thm: Second Asymptotic Log-Optimality of Buy and Hold}, we have $g_n(K) \to g_1^*$ as $n \to \infty$. Therefore, the market portfolio constructed is asymptotically log-optimal. \qedhere
\end{proof}

\medskip
\subsection{Illustrative Examples (Well-Diversified Portfolio with Unequal Weights)}\label{ex:Overconfident Buy and Holder}
Consider a stock market that contains $m\geq 2$ assets and one of them is dominant, say it is the~$j$th asset, but we assume that this information is not revealed to any trader. Now, suppose that a  trader enters the market and incorrectly estimates that the dominant asset to be Asset~$i$, which is different from the truly dominant one.
Based on the estimation, the trader decided to invest a majority of his wealth in the $i$th asset; i.e., $K_i := 1-\varepsilon$ for sufficiently small $\varepsilon \in (0,1)$. However, to hedge the estimation error, he distributed the remaining fraction with equal weights on the rest of the assets; i.e., for each asset $j\neq i$, we put
$
K_j := \varepsilon/(m-1).
$
That is, a market portfolio with unequal weights is constructed by the trader.
Then, according to Theorem~\ref{thm: Second Asymptotic Log-Optimality of Buy and Hold} above, the trader still enjoys the property that $\lim_{n\to \infty} g_n(K) = g_1^* $ with a rate of convergence
$$
\frac{1}{n} \left( \log \frac{m-1}{\varepsilon} +1 -\frac{m-1}{\varepsilon} \right) \leq g_1^* - g_n( K ) \leq \frac{1}{n}\log \frac{m-1}{\varepsilon}.
$$
As a second example, in practice, investing in some exchange-traded fund (ETF) which holds a basket of assets including one which is dominant, the theorem and the associated corollary above are in play.

\medskip
\subsection{Sublinear Rate of Convergence}
We now show that the rate of convergence stated in Theorem~\ref{thm: Second Asymptotic Log-Optimality of Buy and Hold} is sublinear. This is summarized in the next corollary  to follow.

\medskip
\begin{corollary}[Sublinear Rate of Convergence]\label{Corollary: Sublinear Rate of Convergence}
	With the same assumptions used in Theorem~\ref{thm: Second Asymptotic Log-Optimality of Buy and Hold}, the bounds for the performance difference $g_1^*-g_n(K)$ converges to zero sublinearly.
\end{corollary}

\smallskip
\begin{proof}{Proof.} The proof is elementary; hence  we only prove that the upper bound for the performance difference \mbox{$g_1^*-g_n(K)$} converges to zero sublinearly. An almost identical proof can be made for the lower bound. Hence we omitted.  
	With $K_j>0$, we let $C:=\log\frac{1}{K_j}$ and take $x_n:=C/{n}$, which forms a sequence of upper bound for the difference $g_1^*-g_n(K)$. Note that sequence $x_n \to 0$ as $n\to\infty$. To see it converges sublinearly, observe that
	\[
	\frac{|x_{n+1}-0|}{|x_n-0|} =	\frac{{1}/({n+1})}{{1}/{n}} = \frac{n}{n+1} \to 1
	\]as $n \to \infty$. Hence, by definition, the rate of convergence is sublinear. \qedhere
\end{proof}


%
%

\bigskip
\section{Improvement of the Asymptotic Log-Optimality Theorem}\label{Improvement of the Asymptotic Log-Optimality Theorem}
In~\cite{hsieh2019impact}, a so-called high-frequency maximality result was proved for a single asset case. Roughly speaking, it indicates that the high-frequency trading is \textit{unbeatable} in the ELG sense  when there are no transaction costs. 
This motivates us to explore the connection between high-frequency maximality and asymptotic log-optimality. 
In this section,
we first state a conjecture regarding the high-frequency maximality in a multi-asset portfolio management scenario. Then, as supporting evidence, we provide a lemma that proves a special case of the conjecture when a certain independence hypothesis for the returns is assumed.
Subsequently, we indicate that the conjecture, if true, enables us to improve our asymptotic log-optimality result obtained in Section~\ref{SECTION: Main Results}. 

\medskip
\subsection{High-Frequency Maximality Conjecture and a Lemma} 
\smallskip
The conjecture regarding high-frequency maximality is stated below. 

\medskip
\begin{conjecture}[High-Frequency Maximality]\label{conj: general high-frequency maximality}
	For the frequency dependent optimization problem defined in Section~\ref*{SECTION: Preliminaries}, then it follows that
	$
	g_n^* \leq g_1^*
	$
	for all $n \geq 1$. That is, high-frequency rebalancing is unbeatable in the ELG~sense.
\end{conjecture}

\medskip
\begin{remark}[Supporting Evidence of the Conjecture] \rm
	One should note that for $n=1$, the statement holds trivially.
	As a theoretical support, in~\cite{hsieh2019impact}, the conjecture was proved for single asset case.  
	In the following lemma, we prove a special case for the conjecture when an additional independence hypothesis is assumed. The proof is relegated to Appendix and may be skipped by readers who are not interested in the
	technical details.
\end{remark}

\medskip
\begin{lemma}[A Version of High-Frequency Maximality]\label{lemma: Generalized High-Frequency Maximality}
	For the frequency dependent optimization problem defined in Section~\ref*{SECTION: Preliminaries}, if returns $\mathcal{X}_n$ and $X(n)$ are independent for $n\geq 1$, then it follows that
	$
	g_n^* \leq g_1^*
	$
	for all $n\geq 1$. 
\end{lemma}

\medskip
\subsection{Improvement of Asymptotic Log-Optimality Result}
If Conjecture~\ref{conj: general high-frequency maximality} holds, we can further improve the bounds obtained in Theorem~\ref{thm: Second Asymptotic Log-Optimality of Buy and Hold}. This is summarized in the next theorem to follow.

\medskip
\begin{theorem}[Improvement of Asymptotic Log-Optimality of Buy and Hold]\label{thm: Improvement of the Asymptotic Log-Optimality of Buy and Hold}
	Let Asset~$j$ be a dominant asset and Conjecture~\ref{conj: general high-frequency maximality} hold. Given any $K \in \mathcal{K}$ with $K_j >0$, then it follows that
	$
	\lim_{n \to \infty}g_n(K)  = g_1^*
	$
	with rate of convergence above governed~by  
	\[
	0 \leq g_1^*-{g_n}\left( K \right) \leq \frac{1}{n}\left( \log \frac{1}{K_j} -    1 + K_j \mathbb{E}\left[\frac{  \mathcal{R}_{n,j}}{K^T\mathcal{R}_n} \right]\right).
	\]
\end{theorem}

\medskip
\begin{proof}{Proof.}  
	We begin by noting that if the upper and lower bounds stated in the theorem hold, then, with $K_j>0$ being fixed,  both of the bounds converge to zero as $n \to \infty$.  Thus, by Squeeze Theorem, it is readily verified that $\lim_{n\to \infty} g_n(K)=g_1^*.$
	To prove the lower bound, we note that since Conjecture~\ref{conj: general high-frequency maximality} holds, we have $g_1^* \geq g_n^*$ for all $n \geq 1$. 
	Hence, it follows that $
	g_1^* \geq g_n(K)
	$ for all~$n$ and all $K \in \mathcal{K}$. This implies that $g_1^* - g_n(K)\geq 0$, which is desired. 
	To prove the upper bound, we proceeds as follows: For notational convenience, we again work with the total return vector 
	$\mathcal{R}_n =\mathcal{X}_n + \textbf{1}
	$ with $i$th component $\mathcal{R}_{n,i} = \mathcal{X}_{n,i}+1$ and $\textbf{1} =[1 \;\; 1 \cdots \; 1]^T \in \mathbb{R}^m$. 
	Note that $\mathcal{R}_{n,i}>0$ for all $i=1,\dots,m$ and all $n\geq 1$. Then, for $K\in \mathcal{K}$ with $K_j \in (0,1)$, we recall equation~(\ref{eq: g_1 minus gn}), which is used in the proof of Theorem~\ref{thm: Second Asymptotic Log-Optimality of Buy and Hold}. That is,
	\begin{align*}
		g_1^* -{g_n}\left( K \right) &=\frac{1}{n}\log \frac{1}{K_j} - \frac{1}{n}\mathbb{E}\left[ {\log \left( {1 + \sum\limits_{i \ne j} {\frac{{{K_i}{\mathcal{R}_{n,i}}}}{{{K_j}{\mathcal{R}_{n,j}}}}} } \right)} \right].
	\end{align*} 
	Using the fact that $\log(1+z)\geq z/(1+z)$ for all $z>-1$, we obtain
	\begin{align*}
		g_1^* -g_n(K) &\leq 	\frac{1}{n}\log (1/K_j) - \frac{1}{n}\mathbb{E}\left[\frac{ \sum\limits_{i \ne j} {\frac{{K_i}{\mathcal{R}_{n,i}}}{{{K_j}{\mathcal{R}_{n,j}}}}}}{{1 + \sum\limits_{i \ne j} {\frac{{K_i}{\mathcal{R}_{n,i}}}{{{K_j}{\mathcal{R}_{n,j}}}}} }} \right]\\
		&=	\frac{1}{n}\log (1/K_j) - \frac{1}{n}\mathbb{E}\left[ \frac{ \sum\limits_{i \ne j} K_i \mathcal{R}_{n,i} }{{K_j \mathcal{R}_{n,j} + \sum\limits_{i \ne j} K_i \mathcal{R}_{n,i} }} \right]\\[1ex]
		&=	\frac{1}{n}\log  \frac{1}{K_j} - \frac{1}{n}\mathbb{E}\left[ \frac{ \sum\limits_{i \ne j} K_i \mathcal{R}_{n,i} }{K^T\mathcal{R}_n} \right]\\[1ex]
		&=	\frac{1}{n}\log  \frac{1}{K_j} - \frac{1}{n}\mathbb{E}\left[ \frac{ \sum\limits_{i \ne j} K_i \mathcal{R}_{n,i} + K_j\mathcal{R}_{n,j} -  K_j\mathcal{R}_{n,j}}{K^T\mathcal{R}_n} \right]\\[1ex]
		&=	\frac{1}{n}\left( \log \frac{1}{K_j} -    1 + \mathbb{E}\left[\frac{  K_j\mathcal{R}_{n,j}}{K^T\mathcal{R}_n} \right]\right)
	\end{align*}
	which is the desired upper bound stated in the theorem.
	%
	To complete the proof, it remains to verify that the quantity on the right-hand side of the last equality is nonnegative. To see this,
	Using the fact that $\log ({1}/{K_j}) \geq 1-K_j + (1+K_j)^2/2$ for all $K_j\in(0,1)$, we have
	\begin{align*}
		\frac{1}{n}\left( \log  \frac{1}{K_j} -    1 + \mathbb{E}\left[\frac{  K_j\mathcal{R}_{n,j}}{K^T\mathcal{R}_n} \right]\right)  &\geq 	
		\frac{1}{n}\left(-K_j + (1+K_j)^2/2  + \mathbb{E}\left[\frac{  K_j\mathcal{R}_{n,j}}{K^T\mathcal{R}_n} \right]\right) \\[1ex]
		&=	\frac{1}{n}\left( \frac{1}{2}+\frac{K_j^2}{2}  + \mathbb{E}\left[\frac{  K_j\mathcal{R}_{n,j}}{K^T\mathcal{R}_n} \right]\right) \\[1ex]
		&\geq 0
	\end{align*}
	where the last inequality holds because the three terms inside the parenthesis are all  nonnegative.
	Thus, 
	the proof is complete. \qedhere
\end{proof}

\medskip
\begin{remark} \rm
	We note that the both of the lower and upper bounds obtained in the theorem above are sharper than that obtained in Theorem~\ref{thm: Second Asymptotic Log-Optimality of Buy and Hold}.
\end{remark}

\medskip
\subsection{When to Rebalance The Portfolio?}
In this subsection, let us consider the following important question: What if a trader is reluctant to buy and hold for a sufficiently long period and decide to rebalance its position at certain time? 
The following theorem provides a way to address this issue.

\medskip
\begin{theorem}[Rebalancing Criterion]\label{thm: Rebalancing Criterion}  
	Suppose that Conjecture~\ref{conj: general high-frequency maximality} holds and Asset~$j$ is dominant for some $j \in \{1,2,\dots,m\}$. Given any $K \in \mathcal{K}$ with $K_j >0$  and
	the  constant $\varepsilon \in (0, \log (1/ K_j)) $,   there exists an integer $n^* \geq 1$ such that for all $n\geq n^*$, we have $0\leq g_1^* -g_{n}(K) \leq \varepsilon.$
\end{theorem}

\medskip
\begin{proof}{Proof.}
	Let  $K\in\mathcal{K}$ and $\varepsilon \in (0, \log (1/K_j)) $ be fixed, by Theorem~\ref{thm: Improvement of the Asymptotic Log-Optimality of Buy and Hold}, we have, 	for any $n\geq 1$, 
	\begin{align*}
		0 \leq g_1^*-{g_n}\left( K \right) 
		&\leq \frac{1}{n}\left( \log \frac{1}{K_j} -    1 + K_j \mathbb{E}\left[\frac{  \mathcal{R}_{n,j}}{K^T\mathcal{R}_n} \right]\right)\\
		& \leq \frac{1}{n}\log \frac{1}{K_j}
	\end{align*}
	where the last inequality holds because  $K_j \mathcal{R}_{n,j} \leq K^T \mathcal{R}_n$ holds with probability one.
	Now, choose
	$$
	n^* : = \ceil{\frac{1}{\varepsilon} \log \frac{1}{K_j}} 
	$$ 
	where $\ceil{z}:=\min\{m\in \mathbb{Z}: z \leq m\}$ is a ceiling function with $\mathbb{Z}$ being the set of all integers and $z\in\mathbb{R}$. Note that $n^*\geq 1$. By the property that $z\leq \ceil{z}$ for any real number $z$, it is readily verified that 
	$ \log \frac{1}{K_j} \leq \varepsilon n^*.
	$ 
	Now, observe that for any $n \geq n^*$,
	\begin{align*}
		0 \leq g_1^*-{g_{n}}\left( K \right) &\leq \frac{1}{n }\log \frac{1}{K_j} \\
		& \leq \frac{1}{n } n^*\varepsilon \\
		& \leq \varepsilon
	\end{align*}
	and the proof is complete.
	\qedhere
\end{proof}

\medskip
\subsection{Illustrative Example Using Historical Intraday Tick Data}\label{Subsection:Illustrative Example Using Historical Intraday Tick Data}
\begin{figure}[h!]
	\centering
	\includegraphics[width=0.6\linewidth]{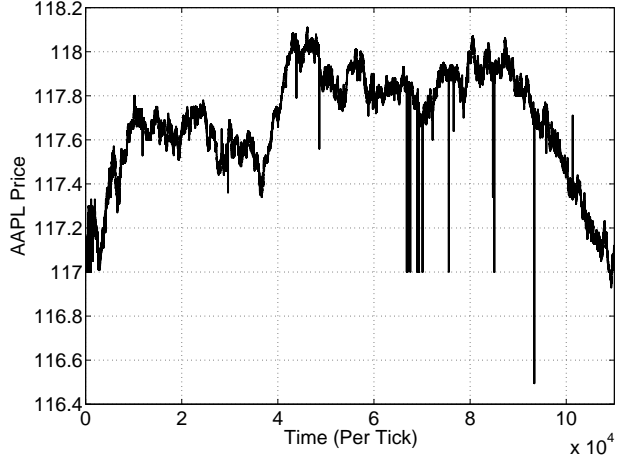}
	\caption{APPLE Stock Prices}
	\label{fig:figaapl151202tick}
\end{figure}

To illustrate Theorem~\ref{thm: Improvement of the Asymptotic Log-Optimality of Buy and Hold}, we consider a stock trading scenario using high-frequency historical intraday data. We assume that there are two traders: One is a high-frequency trader and the other is a buy and holder. Both of them are forming the \textit{same} portfolio with $m=2$ assets where one is risky asset and the other is cash with a constant rate of return $x_1(k):=r = 0$. 
For risky asset, a historical intraday tick data for Apple~\mbox{(ticker: APPL)} for the period 9:30:00 AM to~2:13:47 PM on December 2, 2015 is used; see Figure~\ref{fig:figaapl151202tick} for the price trajectory. During this period, we have~$110,000$ ticks and each ``tick" corresponds to a transaction which takes the realized stock price from $s(k)$ to $s(k+1)$. The average time between arrivals of ticks is about one-tenth of a~second.

\smallskip
Having obtained the high-frequency stock prices, the corresponding returns and their associated compounding returns can be calculated directly.
In particular, take $s(k)$ be the \textit{realized prices} for the Apple stock, we calculate the associated \textit{realized return}, call it $x(k)$, where $ x_2(k):=\frac{s(k+1)-s(k)}{s(k)}.$ The associated \textit{realized compound returns} are $ \chi_{_{2,n}} := \prod_{k = 0}^{n-1}(1+x_2(k))-1$.  
Let the portfolio weight be $K=[K_1 \;\; K_2]^T$ with $K_1=1-K_2>0$ where $K_1,K_2$ are the weights for cash and for the Apple stock, respectively. 
By carrying out the ELG optimization, we obtain  the optimal ELG benchmark for high-frequency trader to be~$g_1^*\approx 6.71 \times 10^{-9}$.

\begin{figure}[h!]
	\centering
	\includegraphics[width=0.6\linewidth]{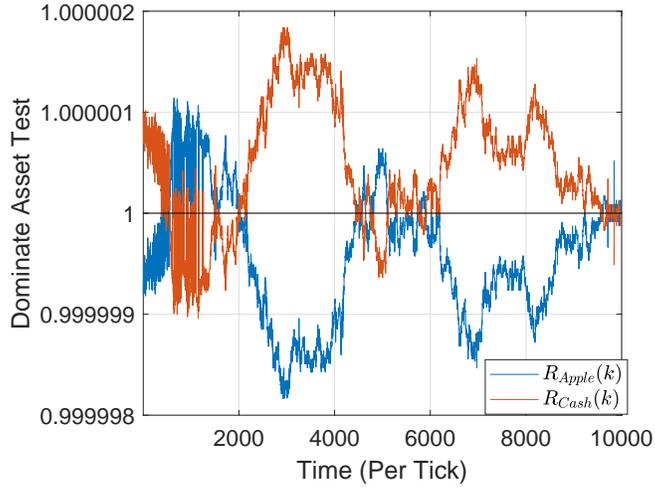}
	\caption{Dominate Asset Test}
	\label{fig:appldominatetest}
\end{figure}

\smallskip
On the other hand, for buy and holder, to apply our theory, we proceed as follows.
We first check the existence of a dominant asset for this two-asset portfolio.
In practice, the realized returns $x_2(k)$ are often not stationary. Thus, to test whether the Apple stock is a dominant asset, we estimate the expected ratio with a sliding window consisting of the most recent~$M$ trading steps; i.e.,
\[
R_{Apple}(k):=\frac{1}{M} \sum_{\ell = 0}^{M-1} \frac{1+r}{1+x(k-\ell)} = \frac{1}{M} \sum_{\ell = 0}^{M-1} \frac{1}{1+x(k-\ell)}.
\]
If $R_{Apple}(k) \leq 1$, then Apple is deemed to be dominant at stage $k$.
Similar test can be made for cash and the associated ratio is denoted by $R_{cash}(k)$.\footnote{To test whether the cash is dominant asset, we estimate  
	$
	R_{cash}(k):=\frac{1}{M} \sum_{\ell = 0}^{M-1} \frac{1+x(k-\ell)}{1+r} .
	$
	If $R_{cash}(k) \leq 1$, then the cash is deemed to be dominant at stage $k$.} In this example, $M=1000$ is used. Figure~\ref{fig:appldominatetest} depicts the ratio for $R_{Apple}$ and $R_{cash}$ for the first $10,000$ ticks. It is typically seen that either the trajectory for $R_{cash}$ or $R_{Apple}$ is below one. Hence, the existence of a dominant asset in the underlying portfolio is assured ``almost surely" except a very few steps where neither of both assets is dominant.

\begin{figure}[h!]
	\centering
	\includegraphics[width=0.6\linewidth]{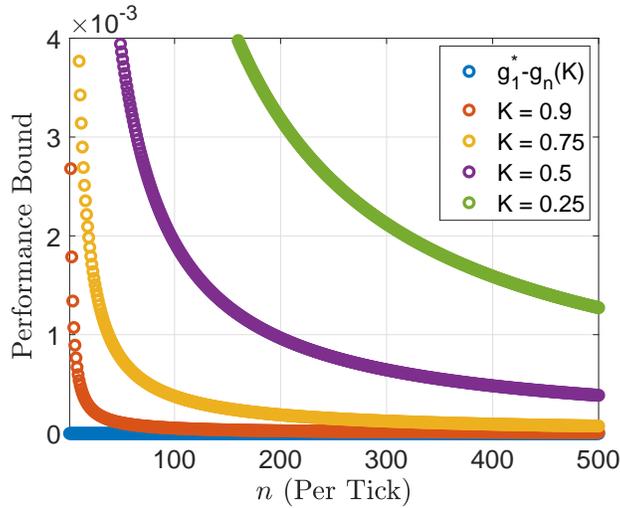}
	\caption{Performance Bounds for Various $K=K_2$.}
	\label{fig:experformaceboundshistoricaldata}
\end{figure}

\smallskip
Having checked the existence of dominant asset, our theory then applies. We are now ready to compare the trading performance via the difference $g_1^*-g_n(K)$ using weights $K=[K_1 \;\; K_2]^T$ with various choices of $K_2$; i.e., $K_2 = 0.25,0.5, 0.75, 0.9$ and $K_1 = 1-K_2$. 
In Figure~\ref{fig:experformaceboundshistoricaldata}, $g_1^*-g_n(K)$, depicted in line with blue color, is upper bounded properly as suggested in Theorem~\ref{thm: Improvement of the Asymptotic Log-Optimality of Buy and Hold} and a convergence pattern is seen as the rebalancing period~$n$~increases.


\smallskip
This example provides a
potential for bridging the theory and practice in stock trading.
Further developments along this line might be fruitful to
pursue as a direction of future research. For example, an systematic way of analysis the existence of dominant asset may be of the next interests to pursue.


\bigskip
\section{Conclusion and Future Work}\label{SECTION: Conclusion}

In this paper, we consider a discrete-time frequency-based portfolio optimization formulation with~$m \geq 2$ assets when the expected logarithmic growth (ELG) rate of wealth is used as the performance metric.
Suppose that the portfolio weights are chosen in a rather \textit{ad-hoc} manner and a buy-and-hold strategy is subsequently used, 
with the aid of the notion of the dominant asset, we showed  that the ELG level obtained by the buy-and-hold strategy tends asymptotically to the optimal ELG level in a sublinear rate of convergence. 
Subsequently, we provided a conjecture regarding the high-frequency maximality. Some theoretical evidence is also provided to support the conjecture. 
This conjecture, if true, enables us to improve the previous log-optimality result.  Also, we proved a result which indicates a way regarding an issue about \textit{when} should one rebalance their~portfolio. 

\smallskip
Regarding future research directions, one natural extension is to study the case where $K_i<0$ is allowed; i.e., \textit{short selling} is considered and the case where $K_i>1$; i.e., leverage is allowed. In addition, it would be of interest to relax some of the assumptions in the formulation from i.i.d. return sequences to time-dependent sequences. Some initial work along this line on a sequential betting problem can be found in \cite{cover2006elements} and \cite{o2020generalization}. In a more general setting, we envision that a stochastic dynamic programming technique may be useful for obtaining the time-varying log-optimal portfolio weights $K_i(k)$. Lastly, we envision that Theorem~\ref{thm: Rebalancing Criterion} may become a useful tool for triggering the signal about when to terminate the trades. Some further studies on developing a trading algorithm may be a promising direction to pursue.

\bigskip

	\section*{Appendix: Proof of Lemma~\ref{lemma: Generalized High-Frequency Maximality}}
	We begin by noting that when $n=1$, the statement holds trivially. Hence, in the sequel, we assume $n>1$.
	Using the shorthand $X_k$ for $X(k) \in \mathbb{R}^m$, for $K \in \mathcal{K}$, 
	we recall  that the  account value of the trader can be expressed as
	$
	V(n,K) 	
	= K^T \mathcal{R}_n V(0)
	$
	where $\mathcal{R}_n := \mathcal{X}_n + \textbf{1} $ is the total return vector with the  $i$th component $\mathcal{R}_{n,i} = \mathcal{X}_{n,i}+1$ and vector $\textbf{1}:=[1 \;\; 1 \cdots 1]^T \in \mathbb{R}^m$.
	To show~$g_n^* \leq g_1^*$, we let $R_n:=X_n + \textbf{1} \in \mathbb{R}^m$ and use the smoothing property of conditional expectation to write the expected logarithmic growth as
	\begin{align}
		g_n(K) 
		&= \frac{1}{n}\mathbb{E}\left[ \log \frac{V(n,K)}{V(0)}\right] \nonumber \\
		&=\frac{1}{n} \mathbb{E}[\log K^T\mathcal{R}_n ]  \label{eq:general gn formula} \\[1ex]
		&=\frac{1}{n} \mathbb{E}\left[\, \mathbb{E}[\log K^T \mathcal{R}_{n} \mid  R_{n-1}]\,\right]\nonumber \\[1ex]
		&= \frac{1}{n}\mathbb{E}\left[\, \mathbb{E}\left[\log\frac{K^T\mathcal{R}_{n}}{K^T R_{n-1}}\bigg| R_{n-1} \right]\,\right] + \frac{1}{n}\mathbb{E}\left[\, \mathbb{E}\left[\log K^T R_{n-1} \mid R_{n-1}\right]\,\right]
		\nonumber \\[1ex]
		& =\frac{1}{n}\mathbb{E}\left[\, \mathbb{E}\left[\log\frac{K^T\mathcal{R}_{n}}{K^T R_{n-1}}\bigg|R_{n-1}\right]\,\right] +\frac{1}{n} \mathbb{E}\left[\log K^TR_{n-1}\right]
		\nonumber \\[1ex]
		& =\frac{1}{n}\mathbb{E}\left[\, \mathbb{E}\left[\log\frac{K^T\mathcal{R}_{n}}{K^T R_{n-1}}\bigg|R_{n-1}\right]\,\right]
		+\frac{1}{n} g_1(K) \label{eq:gn formula}
	\end{align}
	where the last step follows from the fact that the $X_k$ are i.i.d. 
	To simplify the inner conditional expectation above, we use
	the fact that
	\mbox{$
		\mathcal{R}_{n} = R_{n-1} \odot \mathcal{R}_{n-1}  
		$}
	where $\odot$ is the standard Hadamard product; i.e., entry-wise multiplication, with the $i$th component satisfying $\mathcal{R}_{n,i} = R_{n-1,i}\mathcal{R}_{n-1,i}$.
	Now, conditioning on $R_{n-1}=r$ with component $r_i>0$ for all $i=1,\dots,m$, we~observe that
	\begin{align*}
		\mathbb{E}\left[\log\frac{K^T\mathcal{R}_{n}}{K^TR_{n-1}}\bigg|R_{n-1} = r\right]
		& = \mathbb{E}\left[\log\frac{K^T (R_{n-1}\odot \mathcal{R}_{n-1})}{K^TR_{n-1}}\bigg|R_{n-1} = r\right]\\
		& = \mathbb{E}\left[\log\frac{K^T (r\odot \mathcal{R}_{n-1})}{K^Tr}\bigg|R_{n-1} = r\right]\\[1ex]
		& = \mathbb{E}\left[\log\frac{ K^T (D_r  \mathcal{R}_{n-1})}{K^Tr}\bigg|R_{n-1} = r\right] 
	\end{align*}
	where $D_r=diag(r_i) \in \mathbb{R}^{m\times m}$ is the diagonal matrix with vector $r$ as the main diagonals; i.e.,
	\[
	D_r := \begin{bmatrix}
		r_1 & 0 & 0&  \cdots & 0\\
		0 & r_2 & 0 & \cdots &0\\
		0 & \cdots & \ddots & & \vdots\\ 
		\vdots & & &\ddots  & 0\\
		0 & \cdots &\cdots & 0 & r_m
	\end{bmatrix}.
	\] 
	Set 
	$
	K_r^T:=\frac{1 }{K^Tr}K^T  D_r  ,
	$ it is readily verified that $K_r \in \mathcal{K}$. Now, using the hypothesis that $\mathcal{X}_n$ and $X_n$ are independent for all $n\geq 1$, it follows that $\mathcal{R}_n$ and $R_n$ are independent for all $n\geq 1$. Thus, in combination with equation~(\ref{eq:general gn formula}), we obtain
	\begin{align*}
		\mathbb{E}\left[\log K_r^T\mathcal{R}_{n-1} \mid R_{n-1}=r\right]
		&=	\mathbb{E}\left[\log K_r^T\mathcal{R}_{n-1}\right]\\
		&= (n-1)g_{n-1}\left(K_r\right)\\
		&\leq  (n-1)\max_{K\in \mathcal{K}} g_{n-1}(K)\\
		&=(n-1)g_{n-1}^*.
	\end{align*}
	Hence, with $F_{R_{n-1}}$ being the cumulative distribution of $R_{n-1}$, it follows that
	\begin{align*}
		\mathbb{E}\left[\, \mathbb{E}\left[\log\frac{K^T\mathcal{R}_{n}}{K^T R_{n-1}}\bigg|R_{n-1}\right]\,\right]=
		\int \mathbb{E}\left[\log\frac{K^T\mathcal{R}_{n}}{K^TR_{n-1}}\bigg|R_{n-1} = r\right] dF_{R_{n-1}}(r)
		&\leq (n-1)g_{n-1}^*.
	\end{align*}
	In combination of equation~(\ref{eq:gn formula}), we now have
	$
	g_n(K) \le \frac{n-1}{n}g_{n-1}^* + \frac{1}{n}g_1(K).
	$
	Taking the supremum over~\mbox{$K \in \mathcal{K}$} leads to
	\[
	g_{n}^* \leq \frac{n-1}{n}g_{n-1}^* + \frac{1}{n}g_1^*,
	\]
	and, to complete the proof, it is noted that the foregoing argument for~$g_n^*$ also applies to any $g_m^*$ for $m>1$. 
	Hence, 
	$
	g_{m}^* \leq \frac{m-1}{m}g_{m-1}^* + \frac{1}{m}g_1^*.
	$
	Now, with $m=2$, we have $g_2^* \leq g_1^*$. By induction, it is readily shown that $g_n^* \leq g_1^*$ and the proof is complete. \qedhere

\bigskip
\bibliographystyle{siam}
\bibliography{refs.bib}

\end{document}